\newcommand{\useepj}{2}
\ifcase\useepj\or
\documentclass[%preprint,
twocolumn,prl,
               amsmath,10pt]{revtex4}
\usepackage{psfrag,latexsym,epsfig,citee}
\or
\documentclass[11pt]{article}
\usepackage{psfrag,amsmath,epsfig,a4,cite,times}
\fi

\renewcommand{\L}{{\cal L}}
\newcommand{\F}{{\cal F}}

\newcommand{\intx}{\int d^4x}
\newcommand{\VL}{\left( \begin{array}{c}}
\newcommand{\VR}{\end{array} \right)}
\newcommand{\twomat}[1]{\left( \begin{array}{cc} #1 \end{array} \right)}
\def\dg#1{\frac{\delta\Gamma}{\delta#1}}

\def\pslash#1{{\setbox0=\hbox{$#1$}
  \rlap{\ifdim\wd0>.7em\kern.22\wd0\else\kern.1\wd0\fi /}#1}}

\newcommand{\dTB}{\delta t_\beta{}}

\allowdisplaybreaks[1]
\begin{document}
% Title page 
% revtex:
\ifcase\useepj\or
\preprint{DESY--02--068}
\preprint{hep-ph/0205281}
\title{Gauge dependence and renormalization of $\tan\beta$\\
 in the MSSM}
\author{Ayres Freitas}
\author{Dominik St\"ockinger}
\affiliation{Deutsches Elektronen-Synchrotron DESY, 
  \\            D--22603 Hamburg, Germany}
\email{   afreitas@mail.desy.de,
          dominik@mail.desy.de.}
\date{\today}
\begin{abstract}
Well-known and newly developed renormalization schemes for $\tan\beta$ are
analyzed in view of three desirable properties: gauge independence,
process independence, and numerical stability in perturbation theory.
Arguments are provided that no scheme can meet all three requirements,
and as an illustration, a ``No-Go-Theorem'' for the renormalization of
$\tan\beta$ is established. Nevertheless, two particularly attractive
schemes emerge. A discussion about which scheme might be the best
compromise in practice is given.

\end{abstract}

\maketitle
\or
% general
%\begin{titlepage}
\begin{flushright}
DESY--02--068\\
{\tt hep-ph/0205281}\\
\end{flushright}
\vspace{3ex}
\begin{center}
{\Large\bf Gauge dependence and renormalization of $\tan\beta$\\[.2ex]
in the MSSM\\}
\vspace{3ex}
{\large Ayres Freitas and
          Dominik St{\"o}ckinger{\renewcommand{\thefootnote}{\fnsymbol{footnote}}
\footnote{\parbox[t]{10cm}{
          afreitas@mail.desy.de,\\
          dominik@mail.desy.de.}}}}
  \\[2ex]
  \parbox{10cm}{\small\center\em
           Deutsches Elektronen-Synchrotron DESY, 
  \\            D--22603 Hamburg, Germany
  }
\setcounter{footnote}{0}
\end{center}
\vspace{2ex}
\begin{abstract}
Well-known and newly developed renormalization schemes for $\tan\beta$ are
analyzed in view of three desirable properties: gauge independence,
process independence, and numerical stability in perturbation theory.
Arguments are provided that no scheme can meet all three requirements,
and as an illustration, a ``No-Go-Theorem'' for the renormalization of
$\tan\beta$ is established. Nevertheless, two particularly attractive
schemes emerge. A discussion about which scheme might be the best
compromise in practice is given.

\end{abstract}
\centerline
%\end{titlepage}
\fi

\section{Introduction}

The quantity $\tan\beta$ is one of the main input parameters of the
minimal supersymmetric standard model (MSSM). At the tree level it is
defined as the ratio of the two vacuum expectation values $v_{1,2}$ of
the MSSM Higgs doublets,
\begin{align}
\tan\beta & = \frac{v_2}{v_1}.
\label{TBDef}
\end{align}
Owing to its central appearance in spontaneous symmetry breaking,
$\tan\beta$ plays a crucial role in almost all sectors of the MSSM and
has significant impact on most MSSM observables. However, the vacuum
expectation values and $\tan\beta$ are not directly measurable
quantities. In spite of its importance, $\tan\beta$ is an auxiliary
variable. The virtue of $\tan\beta$ is that it can be used as an
easy-to-handle input parameter in terms of which all different
observables can be expressed. In contrast with other parameters, such as the
electron charge or the particle masses, there is no obvious and unique
way to relate $\tan\beta$ to an observable.

The actual definition of $\tan\beta$, its physical meaning and its
relation to observables is given by the choice of a renormalization
scheme. This choice determines the numerical value of $\tan\beta$ as
well as its formal properties, such as gauge dependence and
renormalization-scale dependence.

In the literature, several renormalization schemes for $\tan\beta$
have been proposed and used \cite{Dabelstein,Rosiek,CGGJS96}. Each of
these schemes has specific advantages and disadvantages. In this
paper, we give three criteria that are desirable for a
renormalization scheme for $\tan\beta$:
\begin{itemize}
\item Gauge independence. If the relation between $\tan\beta$ and
observables is gauge independent, the numerical value of
$\tan\beta$ is also gauge independent and a more physical
interpretation of $\tan\beta$ is possible.
\item Process independence. If $\tan\beta$ is defined by a relation to
a physical process, a non-universality and flavor dependence can be
introduced that violates the intuition implied by eq.\ (\ref{TBDef}) that
$\tan\beta$ is a quantity of the MSSM Higgs sector. Process-dependent
schemes have further technical drawbacks like the necessity to
calculate more complicated vertex functions in the determination of
$\delta \tan\beta$ and the possible appearance of infrared divergent
QED or QCD corrections.
\item Numerical stability. The numerical properties of the
renormalization constant
$\delta\tan\beta$ should not spoil the validity of the perturbative
expansion. Generally speaking, the finite contribution to
$\delta\tan\beta$ and the renormalization-scale dependence of
$\tan\beta$ should not be too large.
\end{itemize}
We analyze the properties of the known renormalization schemes in view
of these criteria and devise new schemes in order to satisfy
them. Since it is the most intricate point, we will be concerned with
questions of gauge dependence and gauge independence for the largest
part of the paper.

Our results are negative. We find that the known process-independent
schemes are gauge dependent and that the new gauge-independent schemes
lead to numerical instabilities. Arguments are provided that schemes
satisfying all three criteria do not exist. However, as a result of
our analysis two schemes emerge that are particularly attractive and
useful compromises.

The outline of the present paper is as follows. After briefly
introducing the MSSM Higgs sector in Sec.\ \ref{sec:HiggsSector}, it
is shown in Sec.\ \ref{sec:GaugeDependentSchemes} that the
$\overline{DR}$ scheme as well as the schemes proposed in
\cite{Dabelstein,Rosiek} are gauge dependent already at the one-loop
level. In Sec.\ \ref{sec:GaugeIndSchemes}, three gauge-independent and
process-independent schemes are developed. The numerical instability
induced by these schemes is exhibited in Sec.\ \ref{sec:Numerical}. The
discussion of process-independent schemes is completed in Sec.\
\ref{sec:nogo} by demonstrating that a large class of process- and
gauge-independent schemes leads to numerical instabilities. Finally,
in Sec.\ \ref{sec:processes} the problems caused by process-dependent
schemes are discussed and one useful scheme is presented. Section
\ref{sec:conclusions} contains our conclusions and a discussion of two
attractive renormalization schemes for $\tan\beta$.

\section{The MSSM Higgs sector}
\label{sec:HiggsSector}

The MSSM contains two Higgs doublets
\begin{align}
\label{eq:Higgs}
H_1 &=
\VL v_1 + \frac{1}{\sqrt2}(\phi_1 - i\rho_1) \\[0.5ex] -\phi_1^- \VR, &
H_2 &= \VL \phi_2^+ \\[0.5ex] v_2 + \frac{1}{\sqrt2}(\phi_2 + i\rho_2) \VR ,
\end{align}
whose electrically neutral components are shifted by $v_{1,2}$ in
order to account for the finite vacuum expectation values. The
parameters of the MSSM Higgs sector are
\begin{equation}
g, g', \tilde{m}_1^2, \tilde{m}_2^2, m_3^2,
\label{GaugeHiggsParameters}
\end{equation}
where $g$ and $g'$ are the SU(2) and U(1) gauge couplings,
$\tilde{m}_{1,2}^2=\mu^2+m_{1,2}^2$ with the Higgsino parameter $\mu$,
and $m_{1,2,3}^2$ are soft-breaking parameters. We assume CP
conservation, so all parameters are real. Expressed in terms of
these parameters, the Higgs boson potential reads
\begin{align}
V &= \tilde{m}_1^2 |H_1|^2 + \tilde{m}_2^2 |H_2|^2
    + m_3^2 (H_1^1 H_2^2-H_1^2 H_2^1 + h.c.)
\nonumber\\
 &\quad+\frac{g^2+g'{}^2}{8}(|H_1|^2-|H_2|^2)^2 + \frac{g^2}{2}|H_1^\dagger
    H_2|^2 .
\end{align}

In order to make the physical content more transparent, the parameters
of the gauge and Higgs sector (\ref{GaugeHiggsParameters}) and the shifts
$v_{1,2}$ are reparametrized in terms of
\begin{eqnarray}
e,M_Z,M_W,M_A,\tan\beta,t_1,t_2
\label{NewParameters}
\end{eqnarray}
with ($\tilde{g}^2=g^2+g'{}^2$, $v^2=v_1^2+v_2^2$)
\begin{subequations}
\label{PhysParDef}
\begin{align}
e &= \frac{g g'}{\tilde{g}},\ 
M_Z^2=\frac{\tilde{g}^2 v^2}{2},\ 
M_W^2=\frac{g^2 v^2}{2}, \ 
\tan\beta =\frac{v_2}{v_1},\\
\label{MADef}
M_{A}^2&=\sin^2\beta\frac{t_1}{\sqrt2 v_1}
+\cos^2\beta\frac{t_2}{\sqrt2 v_2}-m_3^2(\tan\beta+\cot\beta),\\
t_1 &= \sqrt2\bigg[\tilde{m}_1^2 v_1+m_3^2 v_2
+\frac{\tilde{g}^2}{4}v_1(v_1^2-v_2^2)\bigg],\\
t_2 &= \sqrt2\bigg[\tilde{m}_2^2 v_2+m_3^2 v_1
-\frac{\tilde{g}^2}{4}v_2(v_1^2-v_2^2)\bigg].
\end{align}
\end{subequations}
% \begin{align}
% e &= \frac{g g'}{\sqrt{g^2+g'{}^2}},\\
% M_Z^2&=\frac{1}{2}(g^2+g^{\prime 2})(v_1^2+v_2^2),\\
% M_W^2&=\frac{1}{2}g^2(v_1^2+v_2^2), \\
% \label{MADef}
% M_{A}^2&=\sin^2\beta\frac{t_1}{\sqrt2 v_1}
% +\cos^2\beta\frac{t_2}{\sqrt2 v_2}-m_3^2(\tan\beta+\cot\beta),\\
% \tan \beta&=\frac{v_2}{v_1},\\
% t_1 &= \sqrt2\bigg[\tilde{m}_1^2 v_1+m_3^2 v_2
% +\frac{g^2+g'{}^2}{4}v_1(v_1^2-v_2^2)\bigg],\\
% t_2 &= \sqrt2\bigg[\tilde{m}_2^2 v_2+m_3^2 v_1
% -\frac{g^2+g'{}^2}{4}v_2(v_1^2-v_2^2)\bigg].
% \end{align}
For vanishing tadpole parameters $t_1=t_2=0$, $M_A^2$ is the mass of
the pseudoscalar $A^0$, and the tree-level Higgs mass eigenstates
are given by the fields $H,h,G^0,A^0,G^\pm,H^\pm$ with
\begin{align}
\VL \phi_1 \\ \phi_2 \VR &=
\twomat{c_\alpha & -s_\alpha \\ s_\alpha & c_\alpha}
\VL H \\ h \VR,\\
\VL \rho_1 \\ \rho_2 \VR &=
\twomat{c_\beta & -s_\beta \\ s_\beta & c_\beta}
 \VL G^0 \\ A^0 \VR,\\
\VL \phi_1^{\pm} \\ \phi_2^{\pm} \VR &=
\twomat{c_\beta & -s_\beta \\ s_\beta & c_\beta}
 \VL G^{\pm} \\ H^{\pm} \VR
\end{align}
and
\begin{align}
\tan 2 \alpha &= \tan 2 \beta \frac{M_{A}^2+M_Z^2}{M_{A}^2-M_Z^2},\quad
-\pi/2<\alpha<0.
\end{align}
Here and in the following we use the abbreviations
$t_\beta=\tan\beta,s_\beta=\sin\beta,c_\beta=\cos\beta$, etc. 

The tree-level Higgs-boson masses read
\begin{align}
\label{eq:MHpmDef}
M_{H^\pm}^{2(0)}&=M_{A}^2+M_W^2,\\
M_{H,h}^{2(0)}&=\frac{1}{2}\left(M_{A}^2+M_Z^2\pm 
\sqrt{(M_{A}^2+M_Z^2)^2 - 4M_Z^2 M_{A}^2 \cos^2(2\beta)}\right).
\end{align}
An important equation is given by
\begin{align}
\cos^2({2\beta}) &= \frac{M_h^2{}^{(0)} M_H^2{}^{(0)}} 
                {M_A^2(M_h^2{}^{(0)} + M_H^2{}^{(0)} - M_A^2)}.
\label{C2BFormula}
\end{align}
It relates $\tan\beta$ to the physical Higgs-boson masses and will be useful
for later purposes.

The counterterms to the MSSM Higgs sector are generated by a
multiplicative renormalization transformation of the parameters and
fields. For the purpose of the present paper, the explicit form of
the field renormalization is not important, but the parameter
renormalization is given by
\begin{align}
g^{(\prime)}&\to g^{(\prime)} + \delta g^{(\prime)},&
\tilde{m}_{1,2}^2 &\to \tilde{m}_{1,2}^2 +\delta \tilde{m}_{1,2}^2
,\nonumber\\
{m}_{3}^2 &\to {m}_{3}^2 +\delta {m}_{3}^2,&
v_{1,2} &\to v_{1,2} +\delta \tilde{v}_{1,2}.
\end{align}
Through the relations
(\ref{PhysParDef}) also renormalization constants $\delta e$, $\delta
M_{Z,W,A}$, $\dTB$ and $\delta t_{1,2}$ are defined, in
particular $\dTB=\tan\beta(\frac{\delta
\tilde{v}_2}{v_2}-\frac{\delta\tilde{v}_1 }{v_1})$.
There are two important relations between $\dTB$ and other
renormalization constants (for $t_1=t_2=0$) that will be of use later:
\begin{align}
\label{TadpoleTBRelation}
\frac{\delta t_1}{v_1}+\frac{\delta t_1}{v_2} &=
\sqrt2(\delta\tilde{m}_1^2 + \delta\tilde{m}_2^2)
 + \sqrt2 \delta m_3^2 (t_\beta+{\textstyle\frac{1}{t_\beta}})
 + \sqrt2 m_3^2 (1-{\textstyle\frac{1}{t^2_\beta}})\delta t_\beta,\\
\delta M_A^2 & = -\delta m_3^2(t_\beta+{\textstyle\frac{1}{t_\beta}})
- m_3^2(1-{\textstyle\frac{1}{t^2_\beta}})\delta t_\beta
+s^2_\beta\frac{\delta t_1}{\sqrt2 v_1}
+c^2_\beta\frac{\delta t_2}{\sqrt2 v_2}.
\end{align}

\section{Gauge dependence of some schemes for $\tan\beta$}
\label{sec:GaugeDependentSchemes}

\subsection{Definition of the schemes}

The actual definition of the parameters (\ref{NewParameters}), their
physical meaning and their relations to experimental quantities is
given by the
choice of a renormalization scheme. In gauge theories a gauge fixing
is necessary for quantization. Since the gauge fixing is unphysical,
the relations between observable quantities do not depend on it, but
the relations between observables and the parameters
(\ref{NewParameters}) can be gauge dependent. Accordingly, in such a
case the values for the parameters extracted from experiment are gauge
dependent.

For $e,M_{Z,W,A}$, the on-shell renormalization scheme (see
\cite{MSSM} for the case of the MSSM)
provides a gauge-independent definition \cite{GaGr99}: $e$ is related
to the effective charge in the Thomson limit, $M_{Z,W,A}$ are the
masses of the $Z,W,A^0$ bosons (defined as the real parts of the poles
of the respective propagators). In addition, setting the renormalized
tadpoles to zero,
\begin{align}
\Gamma_{\phi_1}=\Gamma_{\phi_2}\stackrel{!}{=}0,
\label{TadpoleConditions}
\end{align}
is in agreement with the requirement of gauge independence.
Here and henceforth, $\Gamma$ denotes the generating functional of the
renormalized 1PI vertex functions and
$\Gamma_{\varphi_1\cdots}(p_1,\ldots)$ denotes a vertex function with
incoming fields $\varphi_1\ldots$ and incoming momenta $p_1\ldots$. 

In contrast, for $\tan\beta$ no such gauge-independent
standard-definition is available. In this section we will show
explicitly the gauge dependence of several well-known schemes for
$\tan\beta$. The first of these is the $\overline{DR}$ scheme, which
is defined by the condition
\begin{align}
\overline{DR}:&& \dTB &\stackrel{!}{=} \mbox{pure
divergence},
\label{DRBarCond}
\end{align}
where ``pure divergence'' denotes a term of the order
$\Delta=\frac{2}{4-D}-\gamma_E+\log4\pi$ in dimensional reduction (or
in dimensional regularization when suitable supersymmetry-restoring
counterterms have been added \cite{susyrestoring})
whose prefactor is such that all renormalized quantities are finite.
Further common renormalization schemes for $\tan\beta$ are the ones
introduced by Dabelstein \cite{Dabelstein} and by Chankowski et al.\
\cite{Rosiek} (DCPR). In these schemes one writes $\delta
\tilde{v}_{1,2}=v_{1,2}\delta Z_{1,2}/2 - \delta v_{1,2}$ with the
field renormalization constants $\delta Z_{1,2}$ of the Higgs doublets
and requires the conditions
\begin{align}
\frac{\delta v_1}{v_1} & =  \frac{\delta v_2}{v_2},&
{\rm Re}\hat\Sigma_{A^0 Z}(M_A^2) & =  0,
\end{align}
where the renormalized $A^0Z$ two-point function is decomposed as
$\Gamma_{A^0Z^\mu}(-p,p)= ip_\mu \hat\Sigma_{A^0Z}(p^2)$.
At the one-loop level these conditions lead to
\begin{align}
\mbox{DCPR}:&& \dTB & \stackrel{!}{=}
\frac{1}{2 c_\beta^2 M_Z} {\rm Re}\Sigma_{A^0 Z}(M_A^2)
\label{DabelsteinCond}
\end{align} 
with the unrenormalized $A^0Z$ two-point function $\Sigma_{A^0Z}$. A
similar prescription requiring vanishing $H^+W^-$ mixing instead of
vanishing $A^0Z$ mixing has been used in \cite{Wan01}.

\subsection{Extended Slavnov-Taylor identity as a tool}

In the presented schemes $\tan\beta$ is apparently not directly related to any
observable. It is known that the $\overline{DR}$ scheme leads to a
gauge dependent $\tan\beta$ at the two-loop level \cite{Yamada01}. In
the following, we will show that in fact both 
schemes lead to a gauge dependence of $\tan\beta$ already at the
one-loop level.

The tool we use to determine the gauge dependence is an extended
Slavnov-Taylor identity introduced in \cite{Kluberg,PiSiSTI}:
\begin{align}
\tilde{S}(\Gamma)\equiv S(\Gamma)+ \chi\, \partial_\xi\Gamma & = 0.
\label{eq:ExtSTI}
\end{align}
Here $\xi$ denotes an arbitrary gauge parameter in the
gauge-fixing term and $\chi$ is a fermionic variable acting as the BRS
transformation of $\xi$. $S(\Gamma)$ is the usual Slavnov-Taylor operator
(see Appendix). If not stated otherwise we
leave the form of the gauge-fixing term open. We only assume that the
gauge-fixing term is coupled to auxiliary $B$ fields, 
\begin{align}
\L_{\rm fix} & = \sum_{V=A,Z,W^+,W^-} B^V{}^\dagger \F^V + 
 \frac{\xi}{2}|B^V|^2.
\label{BGauge}
\end{align}
This simplifies the symmetry identities, but for loop calculations,
eq.\ (\ref{BGauge}) is equivalent to $\L_{\rm
fix}=-\frac{1}{2\xi}\sum|\F^V|^2$, which is obtained by eliminating
the $B$ fields using their equations of motion.

In contrast to the usual Slavnov-Taylor identity $S(\Gamma)=0$, the
extended identity (\ref{eq:ExtSTI}) need not be satisfied. If,
however, $\tilde{S}(\Gamma)=0$ holds, then physical quantities ---
expressed as functions of the parameters of the lowest-order
Lagrangian --- are gauge independent. Accordingly,
$\tilde{S}(\Gamma)=0$ implies that the values for the
parameters extracted from experiment are gauge independent. If
suitable $\chi$-dependent terms are added to the action, the extended
Slavnov-Taylor identity holds at the tree level. In
order to satisfy $\tilde{S}(\Gamma)=0$ also at higher orders, the
renormalization conditions must be chosen so as to not contradict
$\tilde{S}(\Gamma)=0$.

We will make use of the extended Slavnov-Taylor identity in particular
in two ways. On the one hand, in our one-loop calculations we employ
regularization by dimensional reduction \cite{Siegel79} assuming all
symmetries are preserved, so $\tilde{S}(\Gamma^{\rm (1),reg})=0$,
where $\Gamma^{\rm reg}$ denotes the unrenormalized vertex functional
and the index ``$^{(1)}$'' denotes the loop order. The identity
$\tilde{S}(\Gamma^{\rm (1),reg})=0$  entails an easy way to calculate
the $\xi$-derivatives $\partial_\xi\Gamma^{\rm (1),reg}_{\varphi_i\cdots}$
of regularized vertex functions.

On the other hand, presuming $\tilde{S}(\Gamma^{\rm reg})=0$ at some
loop order, there is a very simple necessary condition for
$\tilde{S}(\Gamma)=0$ at this order, yielding a simple check whether a
given set of renormalization conditions is compatible with
$\tilde{S}(\Gamma)=0$. As shown in \cite{PiSiSTI,HK95}, the
counterterms in gauge theories can be generally classified into
gauge-dependent total BRS variations and gauge-independent
non-BRS variations. Independent of the
presence of spontaneous symmetry breaking, the non-BRS variations
correspond to the symmetric parameters of the theory, 
whereas the vacuum expectation values as well as field renormalization
constants are related to total BRS variations. This result can
be easily transferred to the case of the MSSM Higgs sector (see also
\cite{MSSM} for the classification of the MSSM counterterms).
It means that in order to satisfy $\tilde{S}(\Gamma)=0$,
\begin{align}
\partial_\xi\delta g^{(\prime)} = \partial_\xi \delta
\tilde{m}_{1,2}^2 = \partial_\xi \delta m_3^2 & = 0
\label{ParGaugeInd}
\end{align}
has to hold.
In contrast, for $\delta \tilde{v}_{1,2}$ and field renormalization constants
a $\xi$ dependence is compatible with $\tilde{S}(\Gamma)=0$. 
Equation (\ref{ParGaugeInd}) also constitutes a sufficient condition for
$\tilde{S}(\Gamma)=0$, 
since if it is satisfied, $\chi$-dependent counterterms can always be
added in such a way that $\tilde{S}(\Gamma)=0$ holds at the considered
loop order.

Thus, eq.\ (\ref{ParGaugeInd}) provides a simple way to check
whether a given renormalization scheme is compatible with
$\tilde{S}(\Gamma)=0$. In the following calculations we will use the
on-shell conditions for $e$, $M_{Z,W,A}$ together with the tadpole
conditions (\ref{TadpoleConditions}). Since these conditions are compatible
with $\tilde{S}(\Gamma)=0$, eq.\ (\ref{ParGaugeInd}) provides in 
particular a check for the renormalization condition for $\tan\beta$
and hence for the gauge independence of $\tan\beta$.

\subsection{Calculating the gauge dependence}

Let us now show the gauge dependence of $\tan\beta$ at the one-loop
level in the $\overline{DR}$ and DCPR schemes
(\ref{DRBarCond}), (\ref{DabelsteinCond}), 
beginning with the $\overline{DR}$ scheme. If $\tilde{S}(\Gamma)=0$ is
to hold, the gauge independence of the counterterms
(\ref{ParGaugeInd}) implies, together with eq.\
(\ref{TadpoleTBRelation}), a certain gauge dependence of the
renormalization constant $\dTB^{\rm fin}$:
\begin{align}
\partial_\xi\left(\frac{\delta t_1}{v_1}+\frac{\delta
t_2}{v_2}\right)^{\rm fin}
& = \left[\sqrt2 m_3^2\left(1- {\textstyle\frac{1}{t_\beta^2}}\right) \right]
\partial_\xi\dTB^{\rm fin}.
\label{DRBarContra}
\end{align}
We only consider the purely finite parts here since the divergent
contributions are restricted by $\tilde{S}(\Gamma^{\rm (1),reg})=0$
and hence are in agreement with $\tilde{S}(\Gamma)=0$.
In the $\overline{DR}$ scheme the finite part of the renormalization
constant $\dTB$ itself is zero and therefore gauge
independent,
\begin{align}
\partial_\xi\dTB^{\rm fin} & = 0,
\label{DRBarContra2}
\end{align}
and the question is whether this is compatible with the gauge
dependence prescribed by eq.\ (\ref{DRBarContra}).

The tadpole counterterms in (\ref{DRBarContra}) are determined by
$\Gamma_{\phi_{1,2}}\stackrel{!}{=}0$ as 
\begin{align}
\delta t_{1,2} & = \Gamma^{\rm (1),reg}_{\phi_{1,2}}.
\end{align}
Using the extended Slavnov-Taylor identity at the regularized level,
$\tilde{S}(\Gamma^{\rm (1),reg})=0$, yields for the l.h.s.\ of
(\ref{DRBarContra}): 
\begin{align}
\partial_{\xi}\left(\frac{\delta t_1}{v_1}+\frac{\delta
t_2}{v_2}\right) & = -\sum_{j=1,2}\Gamma^{\rm (1),reg}_{\chi
Y_{\phi_j}} \sum_{i=1,2}
{\textstyle\frac{1}{v_i}}\Gamma^{(0)}_{\phi_j\phi_i}
\nonumber\\
& = m_3^2\left({\textstyle \frac{1}{v_1^2}-\frac{1}{v_2^2}}\right)
\left(-v_2\Gamma^{\rm (1),reg}_{\chi Y_{\phi_1}}
+v_1\Gamma^{\rm (1),reg}_{\chi Y_{\phi_2}}\right),
\label{DXiTadpoles}
\end{align}
where the $Y_{\phi_i}$ denote the sources of the BRS
transformations of $\phi_i$ used in the Slavnov-Taylor identity.
The $\xi$ dependence of $\Gamma^{\rm (1),reg}_{\phi_{i}}$ and the
results for $\Gamma^{\rm (1),reg}_{\chi Y_{\phi_i}}$ depend on the
specific choice of the gauge fixing. In the $R_\xi$ gauge we obtain
(for the relevant Feynman rules see the Appendix):
\begin{align}
\Gamma^{\rm (1),reg}_{\chi Y_{\phi_i}} & = 
\left(\begin{array}{c}\cos\beta\\\sin\beta\end{array}\right) 
\frac{1}{16\pi^2}\bigg(
\frac{\tilde{g}M_Z}{4}B_0(0,\xi M_Z^2,\xi M_Z^2)
\nonumber\\&
+\frac{g M_W}{2}B_0(0,\xi M_W^2,\xi M_W^2)\bigg),
\label{ResRxi}
\end{align}
where $B_0$ denotes the usual two-point function.
Hence, in the context of the $R_\xi$ gauge, 
\begin{align}
\partial_\xi\left(\frac{\delta t_1}{v_1}+\frac{\delta
t_2}{v_2}\right)^{\rm fin} & = 0, \label{RxiResult}
\end{align}
and the $\overline{DR}$ scheme
is compatible with eqs.\ (\ref{DRBarContra}), (\ref{DRBarContra2}) and hence
with $\tilde{S}(\Gamma)=0$ at the one-loop
level. This confirms the result of \cite{Yamada01}, where one-loop
gauge independence, but two-loop gauge dependence of $\tan\beta$ was
found in the $R_\xi$ gauge. 

\begin{figure}
\begin{center}
\begin{picture}(330,50)
\epsfxsize=20cm
\put(-65,-650){\epsfbox{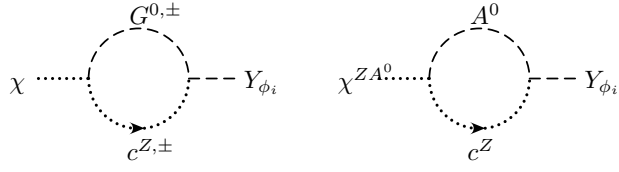}}
\end{picture}
\end{center}
\caption{The one-loop diagrams to the vertex functions 
$\Gamma_{\chi Y_{\phi_i}}^{\rm \,(1),reg}$ and $\Gamma_{\chi^{ZA^0}
Y_{\phi_i}}^{\rm \,(1),reg}$.}
\label{DiagramsDRBar}
\end{figure}
However, in more general gauges (\ref{RxiResult}) is not true. 
As a simple example we consider an infinitesimal deviation from
the $R_\xi$ gauge by introducing a second, infinitesimal gauge
parameter $\zeta^{ZA^0}$ that enters the gauge-fixing function for the
$Z$ boson as follows:
\begin{eqnarray}
\label{GeneralizedGauge}
\F^Z & = & \partial_\mu Z^\mu + M_Z(\xi G^0 + \zeta^{ZA^0}A^0).
\end{eqnarray}
We can study the dependence on this second gauge parameter in the same
way as the dependence on $\xi$. We introduce the variable
$\chi^{ZA^0}$ and consider the Slavnov-Taylor
identity $\tilde{S}(\Gamma)\equiv
S(\Gamma)+\chi^{ZA^0}\partial_{\zeta^{ZA^0}}\Gamma=0$. 
As in the general case, the on-shell conditions for $e$, $M_{Z,W,A}$
and the tadpole conditions (\ref{TadpoleConditions}) are in agreement
with $\tilde{S}(\Gamma)=0$.\footnote{
Clearly, (\ref{GeneralizedGauge}) induces a gauge-dependent mixing of
$A^0$ with $G^0$, $Z^\mu$ and $B^Z$. One might wonder whether this
mixing has an effect on the on-shell condition for $M_A$ and whether
it is still in agreement with the requirement of gauge independence.
As shown in \cite{MSSM}, as long as the auxiliary $B$ fields are not
eliminated from the gauge fixing (\ref{BGauge}), this on-shell
condition can be written as 
$${\rm det}\left(
      \begin{array}{cc}\Gamma_{G^0G^0} & \Gamma_{G^0A^0}\\
                       \Gamma_{A^0G^0} & \Gamma_{A^0A^0}
\end{array}\right)(p^2={\cal M}_A^2)=0$$
with ${\rm Re \cal M}_A^2=M_A^2$. Moreover, with uneliminated $B$
fields there is no contribution of the gauge fixing to
$\Gamma_{G^0G^0}$, $\Gamma_{G^0A^0}$, $\Gamma_{A^0A^0}$, and the
evaluation of the 
renormalization conditions at the tree level is unchanged compared to
the $R_\xi$ case. At higher orders, as shown in \cite{MSSM}, 
this condition for $M_A$ indeed corresponds to the (gauge-independent)
pole of the
$A^0$ propagator, even though the $A^0$ field mixes with $B^Z$ at the
tree level --- or
with $G^0$ and $Z^\mu$ after the elimination of $B^Z$.}
Using this identity we obtain an identity prescribing the
$\zeta^{ZA^0}$ dependence of $\dTB^{\rm fin}$, 
analogously to eq.\ (\ref{DRBarContra}),
\begin{eqnarray}
\partial_{\zeta^{ZA^0}}\left(\frac{\delta t_1}{v_1}+\frac{\delta
t_2}{v_2}\right)^{\rm fin}
& =
\left[\sqrt2 m_3^2\left(1- {\textstyle\frac{1}{t_\beta^2}}\right)
\right]
\partial_{\zeta^{ZA^0}}\dTB^{\rm fin}.
\label{GenContra}
\end{eqnarray}
where, analogously to eq.\ (\ref{DXiTadpoles}), the l.h.s.\ is given by
\begin{align}
\partial_{\zeta^{ZA^0}}\left(\frac{\delta t_1}{v_1}+\frac{\delta
t_2}{v_2}\right) & =m_3^2\left({\textstyle \frac{1}{v_1^2}-\frac{1}{v_2^2}}\right)
\left(-v_2\Gamma^{\rm (1),reg}_{\chi^{ZA^0} Y_{\phi_1}}
+v_1\Gamma^{\rm (1),reg}_{\chi^{ZA^0} Y_{\phi_2}}\right).
\end{align}
However, the one-loop results of the vertex functions involving
$\chi^{ZA^0}$ are (evaluated at $\zeta^{ZA^0}=0$)
\begin{align}
\Gamma^{\rm (1),reg}_{\chi^{ZA^0} Y_{\phi_i}} & = 
\left(\begin{array}{c}-\sin\beta\\\cos\beta\end{array}\right) 
\frac{1}{16\pi^2}\frac{\tilde{g}M_Z}{2}B_0(0,\xi M_Z^2, M_A^2)
\label{ResNonRxi}
.
\end{align}
Hence, the l.h.s.\ of eq.\ (\ref{GenContra}) is non-vanishing:
\begin{align}
\partial_{\zeta^{ZA^0}}\left(\frac{\delta t_1}{v_1}+\frac{\delta
t_2}{v_2}\right)^{\rm fin} & \ne 0. \label{NonRxiResult}
\end{align}
Thus, in the generalized gauge, the $\zeta^{ZA^0}$ independence
of $\tan\beta$ and $\tilde{S}(\Gamma)=0$ imply
\begin{align}
\partial_{\zeta^{ZA^0}}\ \dTB^{\rm fin}
& \ne 0.
\end{align}
This is incompatible with the $\overline{DR}$-condition, which implies
$\partial_{\zeta^{ZA^0}}\dTB^{\rm fin}=0$. 
Conversely, if the $\overline{DR}$ scheme is
used, the relation between $\tan\beta$ and observable quantities,
i.e.\ the value $\tan\beta^{\rm exp}$ extracted from experiment is
gauge dependent:
\begin{align}
\partial_{\zeta^{ZA^0}}\tan\beta_{\overline{DR}}^{\rm exp} &\ne 0.
\end{align}
The different result for eq.\ (\ref{ResNonRxi}) compared to eq.\ (\ref{ResRxi})
can be understood by considering the corresponding Feynman diagrams in
Fig.\ \ref{DiagramsDRBar} (the Feynman rules can be read off from the
Lagrangian given in the Appendix). In the $R_\xi$ gauge fixing only
the Goldstone bosons $G^{0,\pm}$ appear, and correspondingly only
Goldstone bosons contribute as internal scalar lines in the diagrams
for $\Gamma_{\chi Y_{\phi_i}}$. In the generalized gauge, however, we
study the dependence on the $A^0$ part of the gauge fixing, and
correspondingly only $A^0$ appears as an internal scalar line in
$\Gamma_{\chi^{ZA^0}Y_{\phi_i}}$. The couplings of $G^{0,\pm}$ and
$A^0$ to $\binom{Y_{\phi_1}}{Y_{\phi_2}}$ are proportional to
$\binom{\cos\beta}{\sin\beta}$ and
$\binom{-\sin\beta}{\cos\beta}$,
respectively, which explains the results (\ref{ResNonRxi}) and
(\ref{ResRxi}).

\begin{figure}
\begin{center}
\begin{picture}(330,50)
\epsfxsize=20cm
\put(-65,-650){\epsfbox{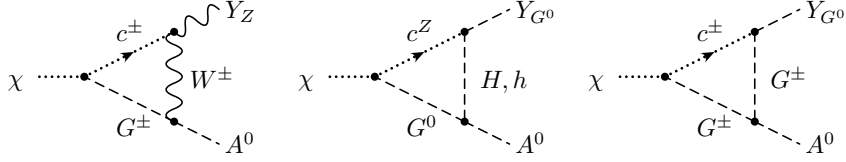}}
\end{picture}
\end{center}
\caption{The one-loop diagrams to the vertex functions 
$\Gamma_{\chi A^0
Y_{Z}}^{\rm \,(1),reg}$ and $\Gamma_{\chi A^0
Y_{G^0}}^{\rm \,(1),reg}$.}
\label{Diagrams}
\end{figure}

In a second step it is quite easy to see the gauge dependence of
$\tan\beta$ in the DCPR schemes defined by eq.\ (\ref{DabelsteinCond}). We know
that the $\overline{DR}$ scheme is in agreement with
$\tilde{S}(\Gamma)=0$ in the $R_\xi$ gauge at the one-loop level. So
we consider the difference
\begin{align}
\partial_\xi(\dTB_{\rm DCPR}-
\dTB_{\overline{DR}}) & \propto \partial_\xi\ {\rm Im}\Gamma_{A^0Z}^{\rm
(1),reg,fin}(M_A^2).
\end{align}
Using $\tilde{S}(\Gamma^{\rm reg})=0$, in the $R_\xi$ gauge the
r.h.s. can be expressed as 
\begin{align}
 -\Big[
\sum_{i=1,2}\Gamma_{\chi Y_{\phi_i}}^{\rm (1),reg}
\Gamma_{A^0 Z \phi_i}^{(0)}+
\sum_{\varphi=Z,G^0}\Gamma_{\chi A^0 Y_{\varphi}}^{\rm (1),reg}
\Gamma_{Z \varphi}^{(0)}\Big]^{\rm fin}\ne0,
\label{NonZero}
\end{align}
which is non-zero as can be easily seen by inspection of the
corresponding one-loop diagrams (see Fig.\ \ref{Diagrams}).
For example, there is an $M_h$-dependent contribution to $\Gamma_{\chi
A^0 Y_{G^0}}$ that cannot be cancelled in eq.\ (\ref{NonZero}).
% \begin{figure}
% \begin{center}
% {\begin{picture}(110,50)
% \psfrag{E1}{\small$\hspace{-0ex}\chi$}
% \psfrag{E2}{\small$Y_Z$}
% \psfrag{E3}{\small$A^0$}
% \psfrag{(1)}{\small$c^\pm$}
% \psfrag{(2)}{\small$W^\pm$}
% \psfrag{(3)}{\small$G^\pm$}
% \epsfxsize=4cm
% \put(-.50,-20.9){\epsfbox{GChiA0YZ.ps}}
% \end{picture} }
% {\begin{picture}(110,50)
% \psfrag{E1}{\small$\hspace{-0ex}\chi$}
% \psfrag{E2}{\small$Y_{G^0}$}
% \psfrag{E3}{\small$A^0$}
% \psfrag{(1)}{\small$c^Z$}
% \psfrag{(2)}{\small$H,h$}
% \psfrag{(3)}{\small$G^0$}
% \epsfxsize=4cm
% \put(-.50,-20.9){\epsfbox{GChiA0YG1.ps}}
% \end{picture} }
% {\begin{picture}(110,50)
% \psfrag{E1}{\small$\hspace{-0ex}\chi$}
% \psfrag{E2}{\small$Y_{G^0}$}
% \psfrag{E3}{\small$A^0$}
% \psfrag{(1)}{\small$c^\pm$}
% \psfrag{(2)}{\small$G^\pm$}
% \psfrag{(3)}{\small$G^\pm$}
% \epsfxsize=4cm
% \put(-.50,-20.9){\epsfbox{GChiA0YG1.ps}}
% \end{picture} }
% \end{center}
% \caption{The one-loop diagrams to the vertex functions 
% $\Gamma_{\chi A^0
% Y_{Z}}^{\rm \,(1),reg}$ and $\Gamma_{\chi A^0
% Y_{G^0}}^{\rm \,(1),reg}$.}
% \label{Diagrams}
% \end{figure}

Hence, in the DCPR schemes $\tan\beta$ is gauge
dependent even at the one-loop level in the $R_\xi$ gauge:
\begin{align}
\partial_{\xi}\tan\beta_{\rm DCPR}^{\rm exp} &\ne 0.
\end{align}

\section{Three gauge-independent schemes for $\tan\beta$}
\label{sec:GaugeIndSchemes}

In the following we present three gauge-independent renormalization
schemes for $\tan\beta$. These schemes share the additional property
that $\tan\beta$ is defined via quantities in the Higgs sector and
without reference to a specific physical process --- in that sense
they are closely related to the tree-level definition
$\tan\beta=\frac{v_2}{v_1}$. The use of these schemes is on the one
hand to demonstrate the existence of such gauge-independent schemes
and on the other hand to illustrate three ways to devise
gauge-independent renormalization schemes.

The first scheme is defined by the requirement that the parameter
$m_3^2$ and its counterterm is gauge independent as dictated by
eq.\ (\ref{ParGaugeInd}), which is most easily 
realized by $\overline{DR}$ renormalization:
\begin{align}
m_3\mbox{ scheme}:&&
\delta m_3^2 & \stackrel{!}{=}  \mbox{pure divergence}.
\label{m3Scheme}
\end{align}
We refer to this scheme as the ``$m_3$ scheme''.
Together with the on-shell conditions for $e,M_{Z,W,A},t_{1,2}$ this
condition fixes the seventh parameter in (\ref{NewParameters}),
$\tan\beta$. Since at the regularized level $\tilde{S}(\Gamma^{\rm
(1),reg})=0$ holds, the divergent part of $\delta m_3^2$ is
gauge independent, hence
\begin{align}
\partial_\xi\delta m_3^2 & = 0
\end{align}
in the $m_3$ scheme, and thus this scheme is
compatible with eq.\ (\ref{ParGaugeInd}) and $\tilde{S}(\Gamma)=0$.
Hence, it defines $\tan\beta$ in  a gauge-independent way.
Using the relation (\ref{MADef}) between $M_A^2$, $\tan\beta$ and
$m_3^2$ we can derive a result for the finite part of $\dTB$:
\begin{eqnarray}
\delta M_A^2{}^{\rm fin} & = & \sin^2\beta\frac{\delta t_1{}^{\rm fin}}{\sqrt2 v_1}
+\cos^2\beta\frac{\delta t_2{}^{\rm fin}}{\sqrt2 v_2}
-m_3^2(1-\cot^2\beta)\dTB{}^{\rm fin}.
\label{tbm3scheme}
\end{eqnarray}
In this way $\dTB$ is expressed via $\delta M_A^2$ and
$\delta t_{1,2}$ by the $A^0$ self energy and the tadpole
contributions. The superscript ``$^{\rm fin}$'' denotes the purely finite
part of the renormalization constants. The result for the purely
divergent part of $\dTB$ is the same as in all other
schemes.

A second gauge-independent scheme can be read off from eq.\
(\ref{DRBarContra}), which is a necessary consequence of
$\tilde{S}(\Gamma)=0$. The most straightforward way to satisfy
eq.\ (\ref{DRBarContra}) is to require
\begin{align}
\mbox{Tadpole scheme}:&&\dTB^{\rm fin} & \stackrel{!}{=}
\frac{1}{\sqrt2 m_3^2(1-\frac{1}{t_\beta^2})}
\left(\frac{\delta t_1}{v_1}+\frac{\delta t_2}{v_2}\right)^{\rm fin}
.
\label{TadpoleScheme}
\end{align}
This scheme will be referred to as the ``Tadpole scheme''. It can also
be obtained in a second, instructive way by temporarily replacing the
tadpole conditions $\Gamma_{\phi_{1,2}}=0$ by the conditions
\begin{align}
\delta t_{1,2}&\stackrel{!}{=}0.
\label{NoTadpoles}
\end{align}
In such a scheme, where the tadpoles are not renormalized, the
renormalization constants $\delta \tilde{v}_{1,2}$ are gauge
independent if $\tilde{S}(\Gamma)=0$ holds (since in this case $\delta
\tilde{v}_{1,2}$ can be entirely expressed in terms of the
gauge-independent constants in eq.\ (\ref{ParGaugeInd})). Hence also
$\dTB$ is gauge independent, and in combination with 
eq.\ (\ref{NoTadpoles}) the $\overline{DR}$-condition
\begin{align}
\dTB^{\rm fin}& \stackrel{!}{=} 0
\label{Tadpole1}
\end{align}
is in agreement with $\tilde{S}(\Gamma)=0$. 

The connected Green functions and the physical content of the theory
do not change by varying $\delta \tilde{v}_{1,2}$ and accordingly
$\delta t_{1,2}$, $\delta t_\beta$, etc.\ while keeping $\delta
g^{(\prime)}$, $\delta\tilde{m}_{1,2}^2$, $\delta m_3^2$ fixed. Equation
(\ref{TadpoleTBRelation}) yields a relation between $\delta
t_{1,2}$, $\delta t_\beta$ and these fixed renormalization constants:
\begin{align}
\dTB^{\rm fin} & =
\frac{1}{\sqrt2 m_3^2(1-\frac{1}{t_\beta^2})}
\left(\frac{\delta t_1}{v_1}+\frac{\delta t_2}{v_2}\right)^{\rm fin}
+ {\cal O}(\delta \tilde{m}_{1,2}^2,\delta m_3^2)
\end{align}
Therefore, for physical quantities the conditions $\Gamma_{\phi_{1,2}}=0$
for the tadpoles and eq.\ (\ref{TadpoleScheme}) for $\tan\beta$
are equivalent to eqs.\ (\ref{NoTadpoles}) and (\ref{Tadpole1}).

By construction, the Tadpole scheme (\ref{TadpoleScheme}) combined
with the on-shell conditions for $e,M_{Z,W,A},t_{1,2}$
is in agreement with the extended Slavnov-Taylor identity
$\tilde{S}(\Gamma)=0$. Hence it provides another gauge-independent
definition of $\tan\beta$. 

In the third scheme, $\tan\beta$ is defined as a combination of
physical Higgs-boson masses in agreement with the lowest-order result
(\ref{C2BFormula}):
\begin{align}
\mbox{HiggsMass scheme}:&&
\cos^2(2\beta) & \stackrel{!}{=} \frac{M_h^2
M_H^2}{M_A^2(M_H^2+M_h^2-M_A^2)} ,
\label{HiggsMassScheme}
\end{align}
where $M_{H,h}^2$ denote the physical masses of $H,h$ (defined as the
real parts of the poles of the respective propagators). This scheme is
denoted as ``HiggsMass scheme''. Since the physical masses are
gauge-independent quantities, this scheme provides manifestly a
gauge-independent definition of $\tan\beta$.

The definition (\ref{HiggsMassScheme}) is problematic, because
due to higher-order corrections the r.h.s.\ can get larger than
unity. In spite of this possibility, evaluating the requirement
(\ref{HiggsMassScheme}) at the one-loop level yields a condition for
$\dTB$ that always has a solution:
\begin{align}
0 & \stackrel{!}{=} \frac{\hat{\Sigma}_{hh}}{M_h^2{}^{(0)}} 
+ \frac{\hat{\Sigma}_{HH}}{M_H^2{}^{(0)}}
- \frac{\hat{\Sigma}_{hh}+\hat{\Sigma}_{HH}}{M_Z^2},
\end{align}
where $\hat\Sigma_{hh,HH}$ denote the renormalized $hh$ and $HH$ self
energies, evaluated on-shell ($p^2=M^2_{h,H}$, respectively). This
equation is linear in $\dTB$ and can therefore always be
solved for $\dTB$ as a function of the unrenormalized self
energies, $\delta M_A^2$ and $\delta t_{1,2}$.

However, the possibility of the r.h.s.\ in eq.\ (\ref{HiggsMassScheme})
getting larger than unity already signals that this HiggsMass scheme
could cause numerical problems. In the next section, the numerical
properties of all three gauge-independent schemes will be discussed.

\section{Numerical instability of the gauge-independent schemes}
\label{sec:Numerical}

Gauge dependence and gauge independence are important conceptual
characteristics of renormalization schemes. But in practice it is also
mandatory that a renormalization scheme does not lead to numerical
instabilities in the quantum corrections to physical processes. The
gauge-dependent schemes presented in section
\ref{sec:GaugeDependentSchemes} have been successfully used in many
practical loop calculations. Only in the calculation of the neutral
Higgs-boson masses \cite{FHHW} slight numerical instabilities caused by the
scheme of \cite{Dabelstein} were reported that are avoided using the
$\overline{DR}$ scheme. In this section we will demonstrate that the
three gauge-independent schemes presented in the foregoing section
lead to much more severe numerical problems at the one-loop level.

A straightforward way to
analyze the numerical behavior of the one-loop corrections is to
derive the renormalization-scale dependence of $\tan\beta$ in the various
schemes. In the DCPR schemes (\ref{DabelsteinCond}) and in the
HiggsMass scheme (\ref{HiggsMassScheme}), $\tan\beta$ is scale
independent because it is defined on-shell, while in the
$\overline{DR}$ scheme, the $m_3$ scheme and the Tadpole scheme the
scale-dependence of $\tan\beta$ is obtained by simply equating
\begin{align}
\left(\tan\beta+\dTB(\bar\mu)\right)_{\rm DCPR}
& = 
\left(\tan\beta(\bar\mu)+\dTB(\bar\mu)\right)_{\rm other\ scheme}
,
\end{align}
where $\bar\mu$ is the renormalization scale used in dimensional
reduction. For the scale-dependence in the $\overline{DR}$ scheme we
obtain a simple analytical formula, which is well-known (see e.g.\
\cite{Yamada01}):
\begin{eqnarray}
\frac{\partial}{\partial\log\bar\mu}\left(\tan\beta(\bar\mu)\right)_{\overline{DR}}
& = & 
\tan\beta\frac{1}{16\pi^2}\left(3h_b^2-3h_t^2+h_\tau^2\right)
,
\end{eqnarray}
where $h_{t,b,\tau}$ are the Yukawa couplings of the top, bottom, and
$\tau$, respectively, and where the contributions of the first two
generations are neglected. The analytical results for the
scale dependence in the other schemes are more complicated, so we restrict
ourselves to a numerical analysis. Table \ref{tab:mudep} contains the
numerical results for the scale dependence for three typical sets of
MSSM-parameters taken from \cite{Scenarios}.

Apparently, the scale dependence of $\tan\beta$ in the $\overline{DR}$ scheme
is quite small, indicating a small uncertainty in one-loop corrections
due to $\dTB$. On the other hand, the scale dependence of
$\tan\beta$ in the $m_3$  and Tadpole schemes can get immensely
large. In practice, for instance changing the scale $\bar\mu$ from
$m_t$ to $M_A$ can cause unacceptably large changes in the numerical
values of one-loop corrections to observables that depend on
$\tan\beta$.

\begin{table}
\begin{center}
\begin{tabular}{|l|r|r|r|r|r|r|}
\hline
 & \multicolumn{3}{|c|}{$\tan\beta=3$} & \multicolumn{3}{|c|}{$\tan\beta=50$} \\
 &  $m_h^{\rm max}$ & large $\mu$ & no mixing
 &  $m_h^{\rm max}$ & large $\mu$ & no mixing \\
\hline
$\overline{DR}$ & $-$0.06 & $-$0.06 & $-$0.06& $-$0.17  & $-$0.17  & $-$0.17\\
$m_3$           & 0.81  & $-$0.46 & $-$0.04& 285.29 & 127.11 & 4.92\\
Tadpole         & 4.50  & $-$0.21 & 1.24 & 370.73 & 140.11 & 34.53\\
\hline
\end{tabular}
\caption{$\partial\tan\beta/\partial{\log\bar\mu}$ in the $\overline{DR}$-
$m_3$, and Tadpole scheme for various parameter scenarios. We have chosen
$M_A=500$GeV, and the remaining parameter values are chosen according
to \cite{Scenarios}.}
\label{tab:mudep}
\end{center}
\end{table}

The fact that $\tan\beta$ is not scale-dependent in the
HiggsMass scheme does not imply that this scheme does not lead to
numerical instabilities. In table \ref{tab:deltaTB} the numerical
values of the purely finite part $\dTB^{\rm fin}$ are
shown in the schemes where it is non-vanishing. Typically, the finite
contribution to a renormalization constant should be suppressed
compared to the respective tree-level parameter. Table
\ref{tab:deltaTB} shows that this is the case only for the
DCPR schemes (and of course for the $\overline{DR}$ scheme), whereas
in the three gauge-independent schemes, in particular in the
HiggsMass scheme,  $\dTB^{\rm fin}$ can exceed $\tan\beta$ by far. The
consequences are very large one-loop corrections to quantities
depending on $\dTB$, signalizing the breakdown of the
validity of the perturbative expansion.

\begin{table}
\begin{center}
\begin{tabular}{|l|r|r|r|r|r|r|}
\hline
 & \multicolumn{3}{|c|}{$\tan\beta=3$} & \multicolumn{3}{|c|}{$\tan\beta=50$} \\
 &  $m_h^{\rm max}$ & large $\mu$ & no mixing
 &  $m_h^{\rm max}$ & large $\mu$ & no mixing \\
\hline
DCPR  & $-$0.10 & $-$0.06 & $-$0.08& 3.56   & 14.47  & 0.46 \\
$m_3$       & 0.56  & $-$0.08 & $-$0.04& 490.45 & $-$67.14 & $-$4.85\\
Tadpole     & 2.64  & $-$0.46 & 0.33 & 624.70 & $-$76.46 & 0.92 \\
HiggsMass     & $-$2.44 & $-$1.83 &$-$1.33 &$-$426.54 &$-$1995.93&$-$314.50\\
\hline
\end{tabular}
\caption{$\dTB^{\rm fin}$ in the DCPR, $m_3$,
Tadpole, and HiggsMass scheme for various parameter scenarios. We have chosen
$\bar{\mu}=M_A=500$GeV, and the remaining parameter values are chosen according
to \cite{Scenarios}.}
\label{tab:deltaTB}
\end{center}
\end{table}

As an example for the influence on the calculation of observables we
consider the one-loop results for the mass of the light CP-even Higgs
boson, $M_h$, according to the strict one-loop formula
\begin{eqnarray}
M_h^2 & = & M_h^2{}^{(0)} - {\rm Re}\hat{\Sigma}_{hh}(M_h^2{}^{(0)}),
\end{eqnarray}
where $\hat{\Sigma}_{hh}$ is the renormalized one-loop $h$-self
energy. The numerical results are presented in table
\ref{tab:HiggsMass} for the case of small $\tan\beta$, where the Higgs-boson
mass is most sensitive to $\tan\beta$.

We observe that for the parameters used here the DCPR schemes and the
$\overline{DR}$ scheme agree well 
numerically, whereas the discrepancy between these schemes and the
gauge-independent ones can be very large. These discrepancies cannot be
interpreted as theoretical errors in the prediction of the Higgs-boson mass
but they are consequences of the invalidity of the perturbative
expansion in the $m_3$, Tadpole and HiggsMass scheme.

In contrast, perturbation theory is trustworthy in the DCPR schemes
and the $\overline{DR}$ scheme. A more detailed 
comparison between the scheme of
\cite{Dabelstein} and the $\overline{DR}$ scheme
taking into account the leading two-loop effects has shown
that the $\overline{DR}$ scheme has generally a better numerical
behavior in certain regions of the parameter space \cite{FHHW}.
\begin{table}
\begin{center}
\begin{tabular}{|l|r|r|r|}
\hline
 & $m_h^{\rm max}$ & large $\mu$ & no mixing \\
\hline
Tree level      &  72.51 & 72.51 & 72.51\\
\hline
DCPR            & 134.44 & 97.40 &112.23\\
\hline
${\overline{DR}}$, $\bar\mu=M_A$ & 135.03  & 97.93 & 112.83\\
${\overline{DR}}$, $\bar\mu=m_t$ & 134.63  & 97.38 & 112.35\\
\hline
HiggsMass       & 119.58  & 81.48 & 102.87\\
\hline
${m_3}$, $\bar\mu=M_A$           & 138.34  & 97.28 & 112.56\\
${m_3}$, $\bar\mu=m_t$           & 143.23  & 93.23 & 112.25\\
\hline
Tadpole, $\bar\mu=M_A$           & 149.94  & 94.08 & 115.18\\
Tadpole, $\bar\mu=m_t$           & 173.45  & 92.21 & 123.97\\
\hline
\end{tabular}
\caption{The light Higgs mass $M_h$ in the five renormalization
schemes and for various parameter scenarios. We have chosen
$\tan\beta=3$ and
$M_A=500$GeV, and the remaining parameter values are chosen according
to \cite{Scenarios}.}
\label{tab:HiggsMass}
\end{center}
\end{table}

\section{Impossibility of defining $\tan\beta$ in the Higgs sector
without introducing  gauge dependence or numerical instability}
\label{sec:nogo}

In the previous section severe numerical problems were found in all
three gauge-independent schemes. In this section it is shown that the
conflict between gauge independence and numerical stability is
unavoidable if $\tan\beta$ is defined via quantities of the Higgs
sector. Since the numerical instability is so strong that the schemes
cannot be used in practice, this result can be interpreted as a
``No-Go''-like theorem for a gauge-independent renormalization of
$\tan\beta$ in the Higgs sector.

More precisely, we study the class of renormalization schemes where
$\dTB^{\rm fin}$ at the one-loop level is given by a linear
combination of finite parts of the on-shell Higgs self energies, their
momentum derivatives $\Sigma'$, and tadpoles: 
\begin{align}
\label{GeneralScheme}
\dTB^{\rm fin} & = \mbox{linear combination of }
\Big(\Sigma_{A^0A^0}(M_A^2), \Sigma_{A^0G^0}(M_A^2), \Sigma_{A^0Z}(M_A^2), 
\Sigma_{HH}(M_H^2),
\nonumber\\
&\quad
\Sigma_{hh}(M_h^2), \Sigma_{Hh}(M_{H,h}^2), \Sigma'_{A^0A^0}(M_A^2), \Sigma'_{HH}(M_H^2),
\Sigma'_{hh}(M_h^2), \delta t_{1}, \delta t_2\Big)^{\rm fin}
\end{align}
The coefficients in this linear combination should be functions of the
parameters of the Higgs potential $e, \tan\beta, M_{Z,W,A}$. The
choice of (\ref{GeneralScheme}) is motivated by the intuition that
$\tan\beta$ is a quantity of the MSSM Higgs sector. All schemes
considered in sections \ref{sec:GaugeDependentSchemes},
\ref{sec:GaugeIndSchemes} belong to this class.

As shown in Appendix \ref{app:nogo}, the most general
gauge-independent schemes of the class (\ref{GeneralScheme}) are given
by
\begin{align}
\dTB^{\rm fin} & =
\dTB^{\rm fin}_{m_3} + 
\Big(a_A K_A + a_H K_H + a_h K_h\Big)^{\rm fin}
\label{GenScheme}
\end{align}
where $a_{A,H,h}$ are coefficients and the quantities $K_{A,H,h}$ are
defined as the following combinations:
\begin{align}
K_A & = \Sigma_{A^0A^0} - 
\sum_{i,j=1,2}\Gamma^{(0)}_{A^0 A^0 \phi_i}
\left(\Gamma^{(0)}_{\phi\phi}\right)^{-1}_{ij}\delta t_{j}
\end{align}
and analogous for $K_{H,h}$. These combinations are gauge independent
as can be seen from the identities (displayed here for the case of
$K_A$)
\begin{align}
\partial_\xi \delta t_i & =
- \Gamma^{\rm (1),reg}_{\chi Y_{\phi_j}} \Gamma^{(0)}_{\phi_j\phi_i}
, &
\partial_\xi \Sigma_{A^0A^0} & =
- \Gamma^{\rm (1),reg}_{\chi Y_{\phi_j}} \Gamma^{(0)}_{A^0A^0\phi_j}
,
\end{align}
derived from $\tilde{S}(\Gamma^{\rm (1),reg})=0$. Besides,
$K_{A,H,h}$ are nothing but the gauge-independent $\delta M_{A,H,h}^2$
mass counterterms in the scheme with $\delta t_{1,2}=0$ (compare
discussion of the Tadpole scheme). 

The numerical instability originates from several terms contributing
to $\dTB$ and the scale dependence
$\partial_{\log\bar\mu}\tan\beta$. In 
particular, in the general scheme (\ref{GenScheme}),
$\partial_{\log\bar\mu}\tan\beta$ contains terms of the order
\begin{align}
\frac{\mu^2}{M_A^2},\quad\frac{M_{\rm Susy}^2}{M_A^2},\quad\frac{M_2^2}{M_A^2},
\label{Large}
\end{align}
which can get large independently of each other.\footnote{In many of
the cases in tab.\ \ref{tab:mudep}, \ref{tab:deltaTB},
\ref{tab:HiggsMass} these terms are actually subdominant due to the
large value of $M_A$. Nevertheless, in general they are very
significant, especially if the ratios in eq.\ (\ref{Large}) get large.}
It is possible to
choose the coefficients $a_{A,H,h}$ in eq.\ (\ref{GenScheme}) such that
these three large terms are exactly cancelled in
$\partial_{\log\bar\mu}\tan\beta$. In this way an optimal, 
gauge-independent scheme with minimal scale dependence is
obtained. Of course, this optimal scheme is nothing but the
HiggsMass scheme (\ref{HiggsMassScheme}), where $\tan\beta$ is defined
on-shell and therefore $\bar\mu$-independent. As seen in section
\ref{sec:Numerical}, the HiggsMass scheme leads to numerically not
acceptable, large contributions to $\dTB$ itself. 

Hence, the only scheme where the contributions (\ref{Large}) are
absent is the numerically inacceptable HiggsMass scheme, and all other
schemes of the form (\ref{GenScheme}) involve the large contributions
(\ref{Large}) to $\partial_{\log\bar\mu}\tan\beta$ and are for this
reason unsuited. None of the gauge-independent schemes
defined in eq.\ (\ref{GenScheme}) can be used in practice.

\section{Process-dependent schemes}
\label{sec:processes}

If $\tan\beta$ is defined via quantities of the Higgs sector there is
an unavoidable conflict between gauge independence and numerical
stability. In order to circumvent these problems one could try to
define $\tan\beta$ outside of the Higgs sector by relating it to a
specific physical process. This method has been adopted in
\cite{CGGJS96}, where it was suggested to use the decay $H^+\to\tau^+
\nu_\tau$. The one-loop corrected decay width to this process reads
\begin{align}
\Gamma[H^+\to\tau^+\nu_\tau] & = \frac{\alpha m_\tau^2 M_{H^\pm}^2
t_\beta^2}{8 M_W^2 s_W^2}\bigg[1+F_{H^+\tau\nu}
\nonumber\\
&\qquad+2\frac{\delta
e}{e}+2\frac{\delta m_\tau}{m_{\tau}} + 2\frac{\delta
t_\beta}{t_\beta} - \frac{\delta M_W^2}{M_W^2} - 2\frac{\delta
s_W}{s_W}\bigg],
\label{Width}
\end{align}
where $F_{H^+\tau\nu}$ is the form factor describing the vertex and external
wave-function corrections to the amplitude $H^+\to\tau^+\nu_\tau$
including the $H^+$--$W^+$ and $H^+$--$G^+$ mixing self energies. 

By requiring that the radiatively corrected decay width retains the
same form as the lowest-order formula, eq.\ (\ref{Width}) can be understood
as a definition of the renormalization constant $\dTB$. As a
consequence of the relation to a physical observable, this definition
of $\tan\beta$ is manifestly gauge independent.

However, this scheme also has several drawbacks. At first, for the
computation of $\dTB$ it is necessary to compute loop corrections to
the three-particle vertex in $F_{H^+\tau\nu}$, which can be difficult beyond
the one-loop level. Furthermore, it is conceptually disadvantageous to
define $\tan\beta$ in a specific process, since in this way it becomes
a non-universal, flavor-dependent quantity.

Finally, the decay vertex $H^+\to\tau^+\nu_\tau$ also receives
QED corrections, which necessarily include contributions with real
photon emission in order to cancel infrared divergences. It is not
possible to separate the QED corrections from the rest of the
electroweak corrections since they are not individually UV finite. For
practical calculations, however, it is unacceptable to include real
bremsstrahlung corrections into the definition of a counterterm, since
this procedure should depend on experimental phase-space cuts and is
therefore technically very involved.

While the first two drawbacks hold for any process-dependent scheme,
the problem posed by the QED corrections can be avoided by the choice
of another process. One possibility is given by the decay
$A^0\to\tau^+\tau^-$. Its one-loop decay width reads
\begin{align}
\Gamma[A^0\to\tau^+\tau^-] & = \frac{\alpha m_\tau^2 M_{A}^2
t_\beta^2}{8 M_W^2 s_W^2}\bigg[1+F_{A^0\tau\tau}
\nonumber\\
&\qquad
+2\frac{\delta
e}{e}+2\frac{\delta m_\tau}{m_{\tau}} + 2\frac{\delta
t_\beta}{t_\beta} - \frac{\delta M_W^2}{M_W^2} - 2\frac{\delta
s_W}{s_W}\bigg].
\label{WidthA0}
\end{align}
The QED corrections to this decay width consist of the photon loop
contributions to $F_{A^0\tau\tau}$ and $\delta m_\tau$, 
\begin{align}
F_{A^0\tau\tau}&=F_{A^0\tau\tau}^{\rm QED}+F_{A^0\tau\tau}^{\rm weak}
,&
\delta m_\tau &= \delta m_\tau^{\rm QED} + \delta m_\tau^{\rm weak}
.
\end{align}
These QED corrections $F_{A^0\tau\tau}^{\rm QED} + 2\frac{\delta
m_\tau^{\rm QED}}{m_\tau}$  form
a UV-finite subset of the full electroweak one-loop corrections. The
reason for this difference to the decay $H^+\to\tau^+\nu_\tau$ is that
the latter process relies substantially on the SU(2) symmetry and
$\gamma$, $Z$ and $W$ loops have to be summed to yield a UV-finite
result. In contrast, the QED corrections to $A^0\to\tau^+\tau^-$ can
be thought of as being generated by an effective theory containing
essentially the $A^0\tau\tau$ vertex and the photon and are therefore
naturally finite.

Owing to the finite QED corrections, a possible definition of
$\tan\beta$ is given by requiring that the pure weak corrections in
eq.\ (\ref{WidthA0}) cancel and that the exact decay width is given by the
tree-level result plus QED corrections:
\begin{align}
&\bigg[1+F_{A^0\tau\tau}^{\rm weak}+2\frac{\delta
e}{e}+2\frac{\delta m_\tau^{\rm weak}}{m_{\tau}} + 2\frac{\delta
t_\beta}{t_\beta} - \frac{\delta M_W^2}{M_W^2} - 2\frac{\delta
s_W}{s_W}\bigg]
 \stackrel{!}{=}
1
\label{A0Def}
,
\\
&\Gamma[A^0\to\tau^+\tau^-]  = \frac{\alpha m_\tau^2 M_{A}^2
t_\beta^2}{8 M_W^2 s_W^2}
\bigg[1+F_{A^0\tau\tau}^{\rm QED}
+2\frac{\delta m_\tau^{\rm QED}}{m_{\tau}}\bigg]
.
\end{align}
With this definition, $\tan\beta$ is gauge independent and not
affected by QED-corrections and infrared divergences. Moreover, the
one-loop correction to the decay $A^0\to\tau^+\tau^-$ using the
renormalization scheme of \cite{Dabelstein} is quite small
\cite{DabelDecays}. Hence, if $\tan\beta$ is defined by eq.\ (\ref{A0Def}),
it does not suffer from the numerical instabilities found for the
cases of the gauge-independent schemes in sections
\ref{sec:Numerical}, \ref{sec:nogo}.

\section{Conclusions}
\label{sec:conclusions}

In this paper, the renormalization of $\tan\beta$ has been studied
in view of the criteria gauge independence, process independence, and
numerical stability. The $\overline{DR}$ scheme as well as the schemes
presented in \cite{Dabelstein,Rosiek} have been shown to imply a
gauge dependence of $\tan\beta$ already at the one-loop
level. Therefore, three gauge-independent schemes have been developed
--- however, using these schemes produces unacceptably large numerical
instabilities in higher-order calculations. The conflict between gauge
independence and numerical stability has been made more explicit by
a general statement about a large class of process-independent schemes, where
$\tan\beta$ is defined via quantities of the Higgs sector. We have
shown that all gauge-independent schemes of this class lead to
numerical instabilities. Hence, it seems to be impossible to find any
renormalization prescription that satisfies all three above criteria.

As a way out of these problems, process-dependent schemes can be used,
but such schemes also have specific drawbacks. Conceptually, the
flavor dependence of $\tan\beta$ is unsatisfactory, and technically,
the necessity to calculate three-point functions and the possible
appearance of QED  and QCD corrections are disadvantages.

In the course of our analysis, two schemes emerge as the best
compromises both in conceptual and practical respects. On the one
hand, the $\overline{DR}$ scheme is the most advantageous among the
process-independent schemes. It is technically very convenient,
numerically perfectly well behaved --- and although it is in general
gauge dependent, it is not gauge dependent at the one-loop level in
the context of the important class of $R_\xi$ gauges. On the other
hand, defining $\tan\beta$ via the decay $A^0\to\tau^+\tau^-$ provides
a particularly attractive process-dependent alternative. In this
scheme, $\tan\beta$ is directly connected to an observable and
therefore gauge independent as well as renormalization-scale
independent. Furthermore, this specific process is theoretically very
clean since it involves no QCD corrections at the one-loop level and
the QED corrections can be split off.

Both the $\overline{DR}$ scheme and the $(A^0\to\tau^+\tau^-)$ scheme
have specific advantages, so depending on the situation one or the
other can be more useful in practice. However, finally a
decision should be made for one definition of $\tan\beta$,
since a common renormalization scheme is important to allow direct
comparisons between different higher-order calculations. In the
$(A^0\to\tau^+\tau^-)$ scheme, the specific process is chosen merely for
technical reasons. From an experimental point of view, the decay
$A^0\to\tau^+\tau^-$ is only one possibility amongst a variety of
potential observables for the determination of $\tan\beta$ (see e.g.\
\cite{TBexp}); a key observable for the definition of $\tan\beta$ does
not exist. This reflects the fact that $\tan\beta$ is an auxiliary
parameter. Accordingly, the advantages of the process-dependent scheme
appear less significant. Owing to its technical convenience and 
its process independence, we assess the $\overline{DR}$ scheme as the
best choice for defining $\tan\beta$.

\paragraph{Acknowledgments:} We are grateful to J.\ Guasch and W.\
Hollik for valuable discussions.

\begin{appendix}

\section{Slavnov-Taylor operator}

The (gauge part of the) Slavnov-Taylor operator of the MSSM can be
written in the form
\begin{align}
S(\Gamma) &=
 \intx\bigg[\sum_{\stackrel{\varphi=\phi_{1,2},G^0,A^0,G^\pm,H^\pm,}{\rm
 other\ fields}}
 \dg{Y_{\varphi}}\dg{\varphi}
\nonumber\\
& \quad + \sum_{V=A,Z,W^+,W^-}\bigg(\dg{Y_{V}^\mu}\dg{V_\mu}
 + \dg{Y_{c^V}}\dg{c^V} + B^V\dg{\bar{c}^V}\bigg)
\bigg],
\end{align}
where ``other fields'' stands for the (s)quark, (s)lepton, chargino
and neutralino fields. For the full form also including supersymmetry
and translational ghosts, see ref.\ \cite{MSSM}. The fields $B^V$ are
auxiliary fields that couple to the gauge-fixing term and are defined
as the BRS transformation of the Faddeev-Popov antighosts
$\bar{c}^V$. The fields $Y_{\varphi_i}$ are sources for the BRS
transformations of the fields $\varphi_i$. For our purposes, the
Feynman rules involving $Y_{Z,G^0,A^0,\phi_{1,2}}$ are important. They
can be derived from the Lagrangian ($s$ denotes the generator of BRS
transformations):
\begin{align}
\L_{Y}  & =  Y_{\phi_1}s\phi_1 + Y_{\phi_2}s\phi_2 + 
Y_{G^0}sG^0 + Y_{A^0}sA^0 + 
Y_{Z^\mu} sZ^\mu + \ldots\nonumber\\
& =  Y_{\phi_1}\frac12\left[g\left(ic^+(c_\beta G^- - s_\beta H^-)+
h.c.\right) + \tilde{g} c^Z(-c_\beta G^0 + s_\beta A^0)\right]
\nonumber\\
&+Y_{\phi_2}\frac12\left[g\left(-ic^+(s_\beta G^+ + c_\beta H^+)+
h.c.\right) + \tilde{g} c^Z(-s_\beta G^0 - c_\beta A^0)\right]
\nonumber\\
&+Y_{G^0}\frac12\left[g\left(c^+(- G^- )+
h.c.\right) + \tilde{g} c^Z(c_\beta \phi_1 + s_\beta \phi_2) +
2 c^Z M_Z\right]
\nonumber\\
&+Y_{A^0}\frac12\left[g\left(c^+(- H^- )+
h.c.\right) + \tilde{g} c^Z(-s_\beta \phi_1 + c_\beta \phi_2) \right]
\nonumber\\
&+Y_{Z^\mu}\left[\partial^\mu c^Z - g c_W(-iW^+
c^-  + iW^- c^+)\right]
+\ldots
\end{align}
with $\tilde{g}=\sqrt{g^2+g'{}^2}$ and $c_W=\frac{M_W}{M_Z}$.

\section{Gauge-fixing terms}

For most parts of the paper we leave the form of the gauge
fixing open. However, in the one-loop calculations two specific
gauge-fixing terms are used. The $R_\xi$ gauge is defined by the
choice
\begin{align}
\F^A   & =  \partial^\mu A_\mu ,\\
\F^Z   & =  \partial^\mu Z_\mu + M_Z \xi G^0,\\
\F^\pm & =  \partial^\mu W^\pm_\mu \pm i M_W\xi G^\pm,\\
\L_{\rm fix} & = \sum_{V=A,Z,W^+,W^-} B^V{}^\dagger \F^V + 
 \frac{\xi}{2}|B^V|^2
\end{align}
for the gauge-fixing functions and the gauge-fixing Lagrangian.
For loop calculations, it is useful to eliminate the $B$ fields
via their equations of motion, yielding $\L_{\rm
fix}=-\frac{1}{2\xi}\sum|\F^V|^2$. 
The $\chi$-dependent terms (as well as the ghost terms) in the
Lagrangian that are necessary to satisfy $\tilde{S}(\Gamma^{(0)})=0$
are obtained from $\L_{\rm
fix,ghost,\chi}=(s+\chi\partial_\xi)(\bar{c}^V(\F^V+\frac{\xi}{2}B^V))$,
where $s$ denotes the BRS operator. For eliminated $B$ fields they read
\begin{align}
\L_\chi^{R_\xi-\rm gauge} & = -\frac{1}{2\xi}\chi
\Big[\bar{c}^A \partial^\mu A_\mu + \bar{c}^Z (\partial^\mu Z_\mu-\xi
M_Z G^0)\nonumber\\
&\quad + \bar{c}\,^-(\partial^\mu W_\mu^+ - i\xi M_W G^+)
+  \bar{c}\,^+(\partial^\mu W_\mu^- + i\xi M_W G^-)\Big].
\end{align}
In the generalized gauge defined by eq.\ (\ref{GeneralizedGauge}), the
correct form of the $\chi^{ZA^0}$-dependent terms is given by
\begin{align}
\L_{\chi^{ZA^0}} & = \chi^{ZA^0} \bar{c}^Z M_Z A^0.
\end{align}

\section{General gauge-independent scheme}
\label{app:nogo}

In this section we show that the most general gauge-independent
renormalization scheme of the class (\ref{GeneralScheme}) is given
by eq.\ (\ref{GenScheme}). 

Since the $m_3$ scheme is gauge independent, any other
gauge-independent scheme has to satisfy
\begin{align}
\partial_\xi(\dTB-(\dTB)_{m_3}) & = 0,
\end{align}
leading us to the question on which combinations of the quantities in
eq.\ (\ref{GeneralScheme}) are gauge independent. In order to answer this
question it is sufficient to study linear combinations of the
quantities in eq.\ (\ref{GeneralScheme}), leaving away
$\Sigma_{A^0A^0,HH,hh}$ and $\Sigma_{A^0Z}$ since the
gauge-independent combinations $K_{A,H,h}$ involving
$\Sigma_{A^0A^0,HH,hh}$ are already known and since $\Sigma_{A^0Z}$
itself is a linear combination of $\Sigma_{A^0G^0}$ and $\delta
t_{1,2}$ as a consequence of the Slavnov-Taylor identity.

The gauge dependence of the remaining quantities can be expressed using
the identity $\tilde{S}(\Gamma^{(1),\rm reg})=0$ as
\begin{align}
&\partial_\xi \left(\delta t_1, \delta t_2, \Sigma_{A^0G^0},
\Sigma_{Hh}(M_{H}^2), \Sigma_{Hh}(M_{h}^2), \Sigma'_{A^0A^0},
\Sigma'_{HH}, \Sigma'_{hh}\right)^T
\nonumber\\
={\cal M}&\left(
\Gamma^{(1),\rm reg}_{\chi Y_{\phi_1}},
\Gamma^{(1),\rm reg}_{\chi Y_{\phi_2}},
\Gamma^{(1),\rm reg}_{\chi A^0 Y_{G^0}}\Gamma^{(0)}_{G^0G^0}
+\Gamma^{(1),\rm reg}_{\chi A^0 Y_{Z^\mu}}\Gamma^{(0)}_{G^0Z^\mu},
\right.
\nonumber\\&\qquad
\Gamma^{(1),\rm reg}_{\chi HY_{h}}(M_H^2),
\Gamma^{(1),\rm reg}_{\chi hY_{H}}(M_h^2),
%\nonumber\\&\qquad\qquad
\left.
\Gamma^{(1),\rm reg}_{\chi A^0Y_{A^0}},
\Gamma^{(1),\rm reg}_{\chi HY_{H}},
\Gamma^{(1),\rm reg}_{\chi hY_{h}}
\right)^T ,
\label{GaugeDepFunctions}
\end{align}
where ${\cal M}$ denotes an invertible matrix whose entries consist of
tree-level expressions like $\Gamma^{(0)}_{\phi_i\phi_j}$
etc.

Calculation of the vertex functions involving $\chi$ shows that
in the $R_\xi$ gauge there is only one vanishing linear combination,
namely $s_\beta\Gamma^{(1),\rm reg}_{\chi Y_{\phi_1}} -
c_\beta\Gamma^{(1),\rm reg}_{\chi Y_{\phi_2}}=0$. The other vertex
functions involving $\chi$ in eq.\ (\ref{GaugeDepFunctions}) are linearly
independent. Correspondingly, the combination (\ref{RxiResult}) is the
only gauge-independent linear combination of the quantities on the
l.h.s.\ of eq.\ (\ref{GaugeDepFunctions}) in the $R_\xi$ gauge. However, in
more general gauges this combination is not gauge independent (see
eq.\ (\ref{NonRxiResult})). Hence, there is no gauge-independent linear
combination of the quantities on the l.h.s.\ of
eq.\ (\ref{GaugeDepFunctions}). 

This confirms that the expressions $K_{A,H,h}$ 
are the only gauge-independent combinations of the quantities in
eq.\ (\ref{GeneralScheme}), as was to be shown.

\end{appendix}

\begin{flushleft}

\end{flushleft}
\end{document}